\documentclass[aps,prb, twocolumn,superscriptaddress,longbibliography]{revtex4-1}
\usepackage{color}
\usepackage{soul}
\usepackage{dcolumn}
\usepackage{graphicx}
\usepackage{amsmath}
\usepackage{amssymb}
\usepackage{bm}

\begin{document}










\title{The multi-terminal Josephson effect}


\author{N. Pankratova}
\affiliation{Department of Physics, Joint Quantum Institute, and Center for Nanophysics and Advanced Materials,
	University of Maryland, College Park, MD 20742, USA.}

\author{H. Lee}
\affiliation{Department of Physics, Joint Quantum Institute, and Center for Nanophysics and Advanced Materials,
	University of Maryland, College Park, MD 20742, USA.}

\author{R. Kuzmin}
\affiliation{Department of Physics, Joint Quantum Institute, and Center for Nanophysics and Advanced Materials,
	University of Maryland, College Park, MD 20742, USA.}

\author{K. Wickramasinghe}
\affiliation{Department of Physics, Joint Quantum Institute, and Center for Nanophysics and Advanced Materials,
	University of Maryland, College Park, MD 20742, USA.}
\affiliation{Center for Quantum Phenomena, Department of Physics, New York University, NY 10003, USA}

\author{W. Mayer}
\affiliation{Center for Quantum Phenomena, Department of Physics, New York University, NY 10003, USA}

\author{J. Yuan}
\affiliation{Center for Quantum Phenomena, Department of Physics, New York University, NY 10003, USA}

\author{M. Vavilov}
\affiliation{Department of Physics, University of Wisconsin-Madison, Madison, Wisconsin 53706, USA}

\author{J. Shabani}
\affiliation{Center for Quantum Phenomena, Department of Physics, New York University, NY 10003, USA}

\author{V. Manucharyan}
\affiliation{Department of Physics, Joint Quantum Institute, and Center for Nanophysics and Advanced Materials,
University of Maryland, College Park, MD 20742, USA.}



\date{\today}

\maketitle

\textbf{Establishment of phase-coherence and a non-dissipative (super)current between two weakly coupled superconductors, known as the Josephson effect~\cite{josephson1962possible}, plays a foundational role in basic physics~\cite{barone1982physics} and applications to metrology~\cite{scherer2012quantum}, precision sensing~\cite{clarke2006squid}, high-speed digital electronics~\cite{likharev1991rsfq}, and quantum computing~\cite{devoret2013superconducting}. The junction ranges from planar insulating oxides to single atoms~\cite{scheer1998signature}, molecules~\cite{winkelmann2009superconductivity}, semiconductor nanowires~\cite{jarillo2006quantum, doh2005tunable}, and generally to any finite-size coherent conductor~\cite{le2008phase}.  Recently, junctions of more than two superconducting terminals gained broad attention in the context of braiding of Majorana fermions in the solid state~\cite{fu2008superconducting, lutchyn2010majorana, oreg2010helical, mourik2012signatures, hyart2013flux, aasen2016milestones} for fault-tolerant quantum computation~\cite{kitaev2003fault}, and accessing physics and topology in dimensions higher than three~\cite{riwar2016multi}. Here we report the first observation of Josephson effect in 3- and 4-terminal junctions, fabricated in a top-down fashion from a  semiconductor/superconductor (InAs/Al) epitaxial two-dimensional heterostructure~\cite{shabani2016two}. 
Due to interactions, the critical current of a $N$-terminal junction becomes the \textit{boundary} of a $(N-1)$-dimensional manifold of simultaneously allowed supercurrents. The measured shapes of such manifolds are explained by the scattering theory of mesoscopic superconductivity, and they can be remarkably sensitive to the junction's symmetry class~\cite{altland1997nonstandard}. Furthermore, we observed a notably high-order (up to 8) multiple Andreev reflections~\cite{blonder1982transition} simultaneously across every terminals pair, which verifies the multi-terminal nature of normal scattering and a high interface quality in our devices. Given the previously shown gate-control of carrier density and evidence of spin-orbit scattering in InAs/Al heterostructures, and device compatibility with other 2D materials, the multi-terminal Josephson effect reported here can become a testbed for physics and applications of topological superconductivity. }



\begin{figure}
	\centering
	\includegraphics[width=\linewidth]{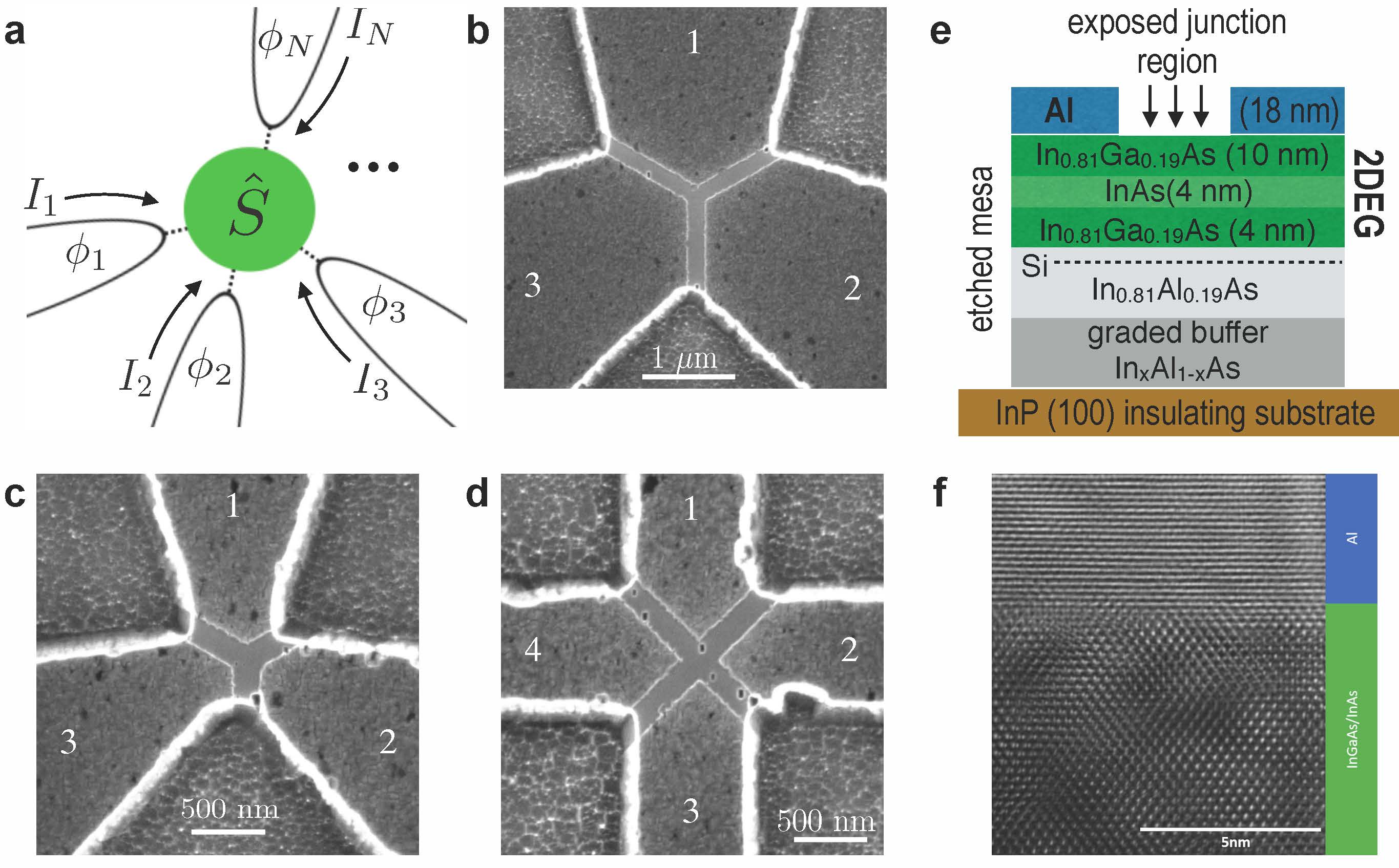}
	\caption{Multi-terminal Josephson junctions based on an epitaxial InAs/Al  heterostructure. (a) A junction of $N$ independent superconductors across a common normal region, characterized by a scattering matrix $\hat S$. Each terminal $j =1,...,N$ has a superconducting phase $\phi_j$ and the associated supercurrent $I_j$. (b-d) Scanning electron microscope images of the nanofabricated junctions of $N = 3$ (b,c) and $N = 4$ (d) copies of which are used in this work. (e) Junction's schematic cross-section revealing the material stack and composition of each layer of the heterostructure. (f) Transmission electron microscope image of the semiconductor/superconductor interface.
	}
	
	\label{fig:Fig1}
\end{figure}

In a conventional tunnel junction considered by Josephson, the supercurrent varies sinusoidally with the difference in phases of the complex-valued order parameter in the two superconducting leads (terminals)~\cite{barone1982physics}. The amplitude of the sinusoid defines the junction's most important parameter -- the critical current. 
A multi-terminal Josephson junction can be conceptually viewed as $N$ independent superconductors coupled through a common element characterized by a scattering matrix $\hat S$ (Fig.~1a). Now a supercurrent in a given terminal periodically depends on all $N-1$ independent phase-differences, which effectively adds an extra dimension to the problem with every new terminal. While this property was considered in the past mostly for device applications~\cite{bevza1979electrically, amin2002dc},  fundamentally new effects in junctions of three and more terminals were recently identified, such as strong breaking of Kramers degeneracy without Zeeman fields~\cite{van2014single} and emergence of Weyl quasiparticle nodes~\cite{riwar2016multi, meyer2017nontrivial, xie2017topological}.
In the simplest manifestation of multi-terminal  Josephson effect, a supercurrent forced into one terminal influences the allowed values of supercurrent in every other terminal. Therefore, traditional measurement of individual critical currents across all possible terminal pairs is insufficient. Instead, one needs to identify all combinations of $N-1$ independent bias currents (the $N$-th one is eliminated by current conservation) for which every terminal maintains zero voltage. For example, the critical current of a 3-terminal junction is a \textit{contour} in a 2D plane.


Junctions with $N>2$ are challenging because the central region must accommodate transparent connections to multiple finite-size leads without exceeding in size the electronic phase-breaking length. In addition, the leads must be relatively equally coupled to each other across the junction. Previous experiments on diffusive metallic systems explored out-of-equilibrium transport~\cite{pfeffer2014subgap} and semi-classical topological aspects of proximity effect~\cite{strambini2016omega}. Notably, two-terminal supercurrents were reported in a nanocross of two crossed-grown InAs nanowires with four superconducting leads~\cite{plissard2013formation}. 
Our junctions are based on an epitaxial heterostructure of III-V materials~\cite{shabani2016two} (Fig.~1b-d). An InAs quantum well is sandwiched between InGaAs barriers to confine a high-mobility 2D electron gas (2DEG) near the surface and simultaneously have a crystalline interface with a superconducting Al layer~\cite{wickramasinghe2018high} (Fig.~1e,f). Three- and four-terminal junctions with sub-micron dimensions were fabricated by selectively removing the unwanted material (METHODS). The exposed 2DEG has the following parameters, obtained through transport measurements on a similar wafer~\cite{wickramasinghe2018high}: Fermi wavelength $\lambda_F \approx 25~\textrm{nm}$, velocity $v_F \approx 10^{6}~\textrm{m/s}$, mean free path $l_e \approx 200~\textrm{nm}$, and phase-breaking length $l_{\phi} \gtrsim 1~\mu m$. We thus expect a quasi-ballistic coherent transport involving a moderate range of $10- 100$ channels.

In the simpler 3-terminal case, the basic characterization consists of grounding one terminal, applying the two DC currents $I_{1}, I_{2}$ to the remaining terminals labeled $1,2$, and simultaneously measuring the two voltages $V_1$ and $V_2$, and differential resistances $dV_{1}/dI_{1}$ and $dV_{2}/dI_{2}$. The data (Fig.~2a,b) contains three generic features. First, the supercurrent state, defined by $V_2 = V_1 =0$, consists of a simply-connected 2D region in the $(I_1, I_2)$-plane. It can be obtained from the data by intersecting the two individual (dark-blue) zero-resistance regions in Fig.~2a and Fig.~2b. The boundary of the supercurrent region defines a \textit{contour} of critical current pairs $\{I_{1}^c, I_{2}^c\}$. The smooth shape of the contour is a clear evidence of interactions between the non-dissipative currents flowing in different terminals, because the fixed value of one strongly influences the maximal value of the other. 
Second, the three ``rays" of reduced differential resistance indicate an out-of-equilibrium condition on currents where only one pair of terminals is under a zero voltage. These rays persists to large currents, at which the voltage across some terminals can exceed the gap of aluminum (Suppl. Mat.). Lastly, the multiple Andreev reflection~\cite{averin1995ac, scheer1998signature} (MAR) of electrons and holes off the superconducting terminals gives rise to a sharp variation of the differential resistance at sub-gap voltages. In the $(I_{1}, I_{2})$-plane, the MAR appears as singularities in the differential resistance aligned along the three zero-voltage rays and around the zero-voltage boundary (Suppl. Mat.). The interference of the high-order dissipative MAR currents in the three terminals creates an intricate conductance pattern immediately outside of the supercurrent state in Fig.~2a,b. 

Up to eight consecutive MAR resonances can be resolved at voltages given by $2\Delta/n$, $n = 1,2,...,8$, where $\Delta$ is interpreted as the superconducting gap induced in the region of the 2DEG directly underneath the aluminum leads~\cite{kjaergaard2017transparent}. The extracted value of the induced gap, $\Delta \approx 175 - 180~\mu \textrm{eV}$, is close to the gap of the Al film~\cite{mayer2018superconducting} ($220~\mu \textrm{eV}$). These observations verify a nearly ballistic transport in the exposed semiconductor junction region and a high transparency boundary with the proximity-superconducting terminals. To our knowledge, such high-order MAR sequences, without missing a single resonance, have never been resolved. The narrower leads junction (Fig.~1c) repeats all the key features of the wider leads one (Fig.~2c,d). The normal state conductance and the supercurrents are reduced roughly proportionally to the width of the leads and the high-order MAR features are smeared. Still, the $n = 1-4$ resonances can be resolved and they give the same value of $\Delta \approx 180~\mu \textrm{eV}$ as the first device. Quite interestingly, the data in Fig.~2 shows no evidence of non-local MAR~\cite{houzet2010multiple} or quartet supercurrents~\cite{cuevas2007voltage,freyn2011production}, recently explored in a InAs nanowire device with three superconducting terminals~\cite{cohen2018nonlocal}.



 

\begin{figure}
	\centering
	\includegraphics[width=\linewidth]{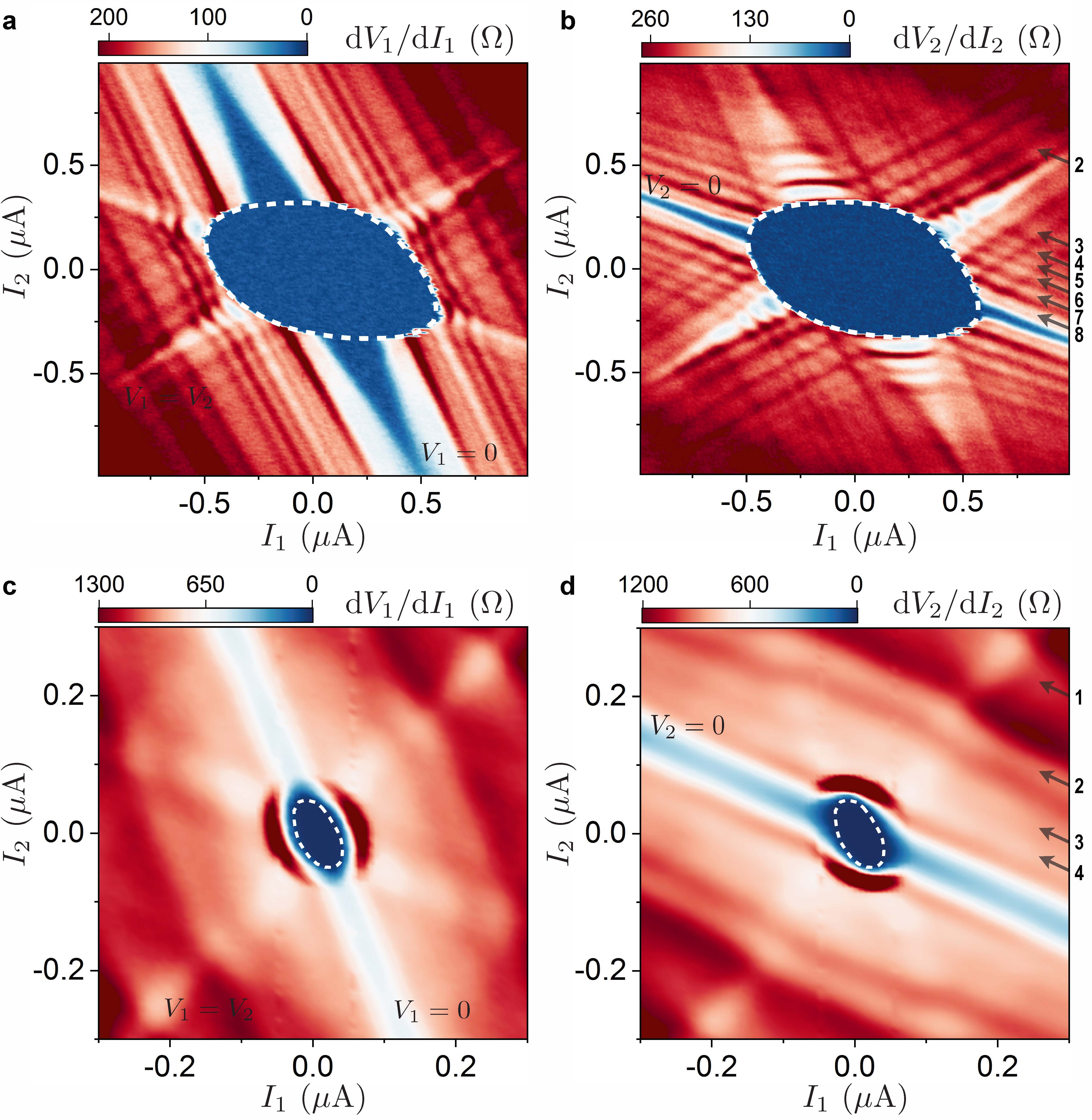}
	\caption{Measurement of 3-terminal critical currents. (a,b) Differential resistance maps of the device shown in Fig.~1b; terminal 3 is grounded. (c,d) Same measurement for the device shown in Fig.~1c. Critical current contours of both devices (dashed white lines) are indicated as a guide for the eye; rays of partial zero-voltage state and the example sequence of multiple Andreev reflection (MAR) between terminals 2 and 3 are labeled. More data details available in Suppl. Mat.
	}
	
	\label{fig:Fig2}
\end{figure} 

\begin{figure*}
	\centering
	\includegraphics[width=0.99\linewidth]{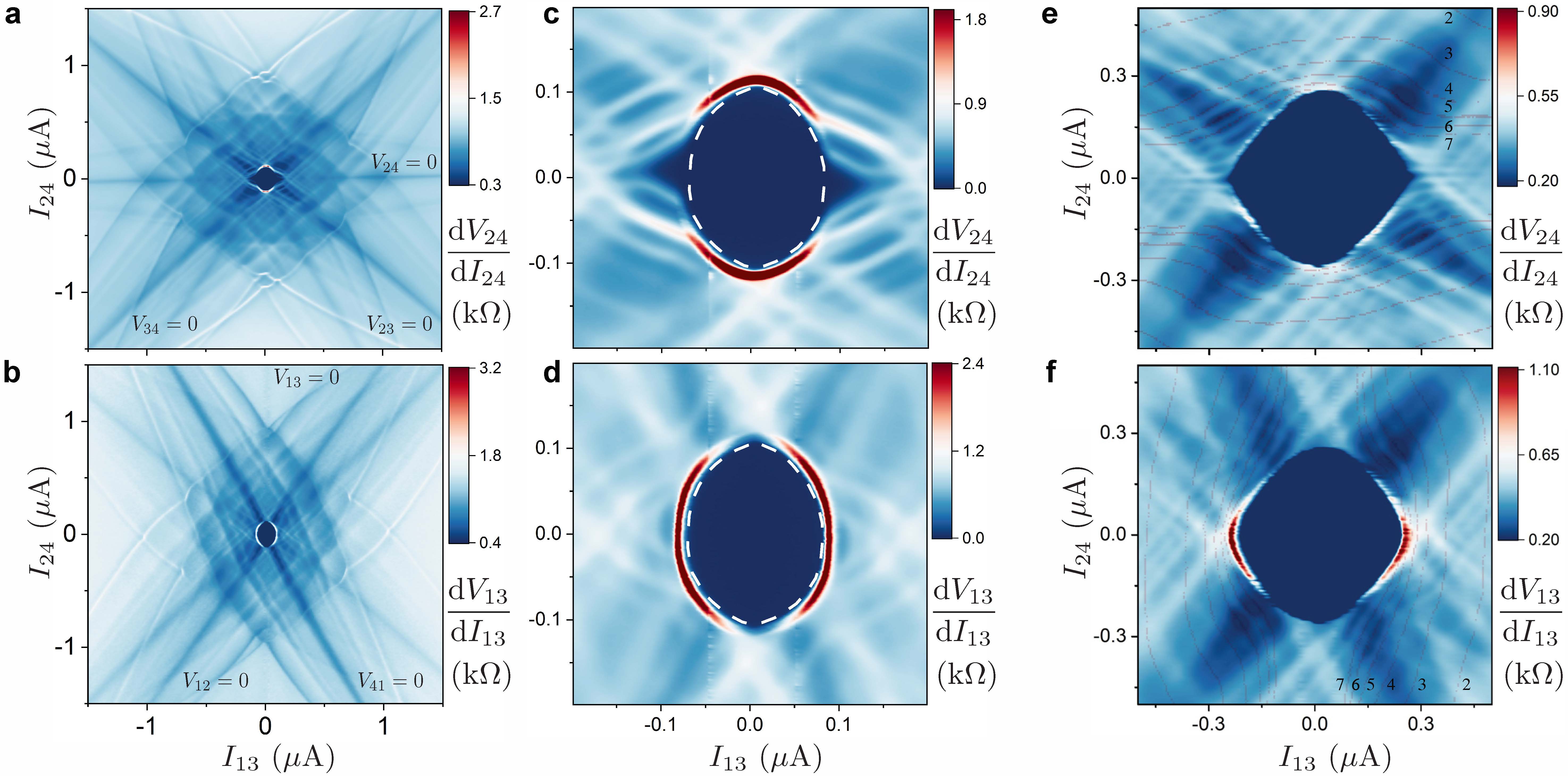}
	\caption{Two-current transport in a symmetrically biased  (see text) 4-terminal device from Fig.~1d. (a,b) Differential resistance across the terminal pairs (2,4) (a) and (1,3) (b). The six radial rays indicate the zero voltage conditions as marked on the graphs. (c,d) Zoom-in on the corresponding supercurrent regions in (a,b). The critical current contour is emphasized by a dashed line. (e,f) Same data as in (a-d) for the same device after thermal cycling. The semi-transparent lines, obtained by measuring voltage, indicate the MAR conditions $V_{24} = 2\Delta/n$ (e) and $V_{13} = 2\Delta/n'$ (f), with $n,n' = 2,3,..., 7$ and $\Delta = 160~\mu\textrm{eV}$.
	}
	
	\label{fig:Fig3}
\end{figure*} 

The measured 3-terminal critical current contours can be understood using the scattering theory of mesoscopic Josephson effect~\cite{beenakker1991universal}. In the absence of a voltage bias, coherent motion of electrons and holes in the semiconductor combined with Andreev reflection off the superconducting terminals gives rise to discrete Andreev bound states (ABS) at energies below the gap. The junction can be viewed as a resonant cavity for coherent electron-hole pairs, controlled by the superconducting phase-differences across the terminals. Each ABS resonance, in general, depends on every possible phase-difference, and it is responsible for carrying supercurrents across the junction. Supercurrents interact because each contributes to shifting the energy of each individual ABS.  The critical current contour can be readily calculated in terms of the normal scattering matrix $\hat S$ of the junction (METHODS). In the absence of microscopic details, $\hat S$ can be chosen randomly from a circular orthogonal ensemble. The orthogonality is required  by the time-reversal symmetry and the energy-independence of matrix elements is justified by the short length of the normal region. 
The main property of contours modeled by orthogonal ensembles is their smooth shape, similarly to what can be seen in Fig.~2. In fact, it was straightforward to find a single matrix $\hat S$ which adequately matched all three critical current contours obtained by permuting the choice of the grounded terminal (Suppl. Mat.).

The critical current of a 4-terminal junction~(Fig.~1d) should in general be a 3D surface. For a simpler data presentation, we switched to a balanced current bias, where terminals are paired and opposite currents are applied to terminals within each pair. For instance, the most intriguing situation is the ``collision" of two supercurrents $I_{24}$ and $I_{13}$ flowing between the oppositely facing even (2,4) and odd (1,3) pairs. 
This way the junction is described by only two independent phase variables and hence the measured four-terminal transport can be interpreted by analogy to the three-terminal data (Fig.~3). Again, the interaction of supercurrents unambiguously manifests itself by the ellipse-like shape of the critical currents contour $\{I_{13}^c, I_{24}^c\}$, approximately aligned along the two current axis (Fig.~3c,d). Outside the critical current contour, there is a region of remarkable out-of-equilibrium coexistence of a non-dissipative current $-10~\textrm{nA} \lesssim I_{24} \lesssim  10~\textrm{nA}$ and a dissipative current $|I_{13}| \approx 100~\textrm{nA}$ flowing directly across each other.
In the resistive state, there are now 6 rays of enhanced conductance associated with a zero voltage across one of the 6 possible terminal pairs (Fig.~3a,b). The high-order MAR between the nearest neighbor terminals dominate the finite-voltage transport in the immediate vicinity of the supercurrent region. 

%


\begin{figure*}
	\centering
	\includegraphics[width=\linewidth]{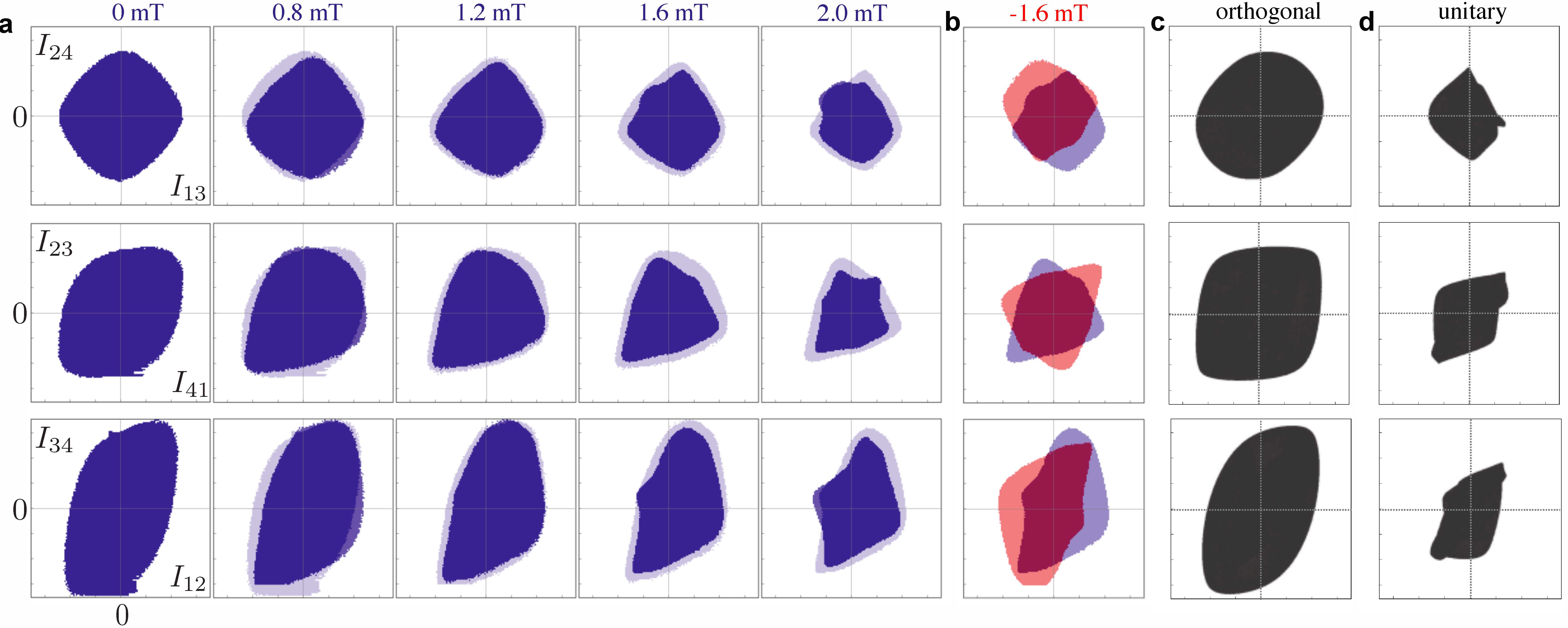}
	\caption{Magnetic field modulation of the critical current contours $\{I_{13}^c, I_{24}^c\}$, $\{I_{41}^c, I_{23}^c\}$, and $\{I_{12}^c, I_{34}^c\}$ of a 4-terminal junction described in Fig~1d and Fig.~3e,f. (a) [Data] the solid blue color indicates the zero-voltage condition on the two bias currents. The copy of the measured shape at a previous field is added in a semi-transparent color for ease of comparison. The range is $100~\textrm{nA/div}$ along both axis. More details available in Suppl. Mat. (b) Data at $B = 1.6~\textrm{mT}$ (transparent blue) vs. $B = -1.6~\textrm{mT}$ (transparent red). Note the symmetry with respect to simultaneous flipping of both axis. (c,d) Scattering theory prediction using orthogonal (c) and unitary (d) ensembles for the matrix $\hat S$.
	}
	
	\label{fig:Fig4}
\end{figure*} 

Thermal cycling to ambient conditions created effectively a new 4-terminal device, which largely reproduced the main features of the original (Fig.~3e,f). The overall scale for conductance and critical currents went up by approximately a factor of two, but the induced gap (determined from MAR) remained unchanged (Suppl. Mat.). Such behavior is expected from the influence of ambient conditions onto the exposed surface 2DEG. 
Importantly, this device clearly shows the high-order MAR resonances between the physically more separated pairs $(1,3)$ and $(2,4)$ (Fig.~3c,d). In fact, the MAR features obeying to $V_{24} = 2\Delta/n$ and $V_{13} = 2\Delta/n'$ ($n, n' = 1,2,...$) appear in the $(I_{13}, I_{24})$-plane only upon crossing the rays $V_{12, 23,34,41} =0$. Moreover, for such conditions, the two MAR currents simultaneously flow across each other through the same physical region in the junction's center. This important observation directly confirms the presence of transparent channels across the semiconductor connecting every pair of terminals.

In the final test, we explore the effect of perpendicular magnetic field $B$ onto all three critical current contours $\{I_{13}^c, I_{24}^c\}$, $\{I_{12}^c, I_{34}^c\}$, and $\{I_{23}^c, I_{41}^c\}$ (Fig.~4a). At $B=0$, it was straightforward to find a matrix $\hat S$ from an orthogonal ensemble which can approximate all three measured contours, similarly to the three-terminal case (Fig.~4c). A magnetic field $B \lesssim 1~\textrm{mT}$ distorts the symmetry of the contours with respect to the simultaneous flipping of both currents. In fact, individual critical currents, obtained by fixing the value of the other current in the pair, can become asymmetric and non-monotonic with field. A qualitatively new feature appears at $B \sim 2~\textrm{mT}$: the contours develop seemingly random small-scale deformations. The corresponding magnetic flux piercing the semiconductor is calculated to be approximately $1/5$ of the superconducting flux quantum, not taking into account flux focusing. We verified that the measured critical current contours are symmetric upon simultaneous flipping of the two currents and the direction of magnetic field (Fig.~4b).

Quite remarkably, these short-scale deformations of the contours can be explained by switching to a unitary ensemble for the scattering matrix $\hat S$, which is appropriate in the absence of time-reversal symmetry inside the junction~(Fig.~4d). In fact, such short-scale contour fluctuations appear to be a generic feature of the unitary ensemble, which furthermore is not significantly sensitive to the number of channels (Suppl. Mat.). This observation lead us to speculate that, unlike the two-terminal currents, which are sensitive only to transmission eigenvalues, the multi-terminal currents can carry qualitatively new information on the symmetry classes of the underlying superconducting Hamiltonians~\cite{altland1997nonstandard}. In particular, it is interesting to identify the effect of spin-rotation symmetry breaking on the critical current contours~\cite{van2014single}. 


In the outlook, the InAs/Al material used in this work has previously demonstrated necessary ingredients for creating solid-state Majorana fermions, such as strong spin-orbit coupling and gate control of the electron density, and possible evidence of topological superconductivity was recently reported~\cite{fornieri2018evidence}. Therefore, besides opening a new experimental direction in mesoscopic superconductivity, our realization of multi-terminal Josephson effect completes an important prerequisite for pursuing the ambitious braiding proposals~\cite{alicea2011non, hyart2013flux, karzig2017scalable}, all based on junctions of multiple topological superconductors through a common coherent conductor. In the applied extreme, the interaction of multi-terminal supercurrents can form a basis for a novel transistor technology for efficient classical computing. 

While preparing this manuscript, we became aware of a recent preprint reporting a multi-terminal superconducting device based on graphene~\cite{draelos2018supercurrent}.

We acknowledge support from NSF-PFC at JQI, NSF-EAGRE, DARPA, and ARO-LPS (NEQST), as well as useful discussions with Alex Levchenko and Manuel Houzet. \\

\noindent\textbf{METHODS}\\
\\
\noindent \textbf{Experimental.} The hybrid quantum well heterostructure (Fig.~1e,f) is grown using a specially developed III-V molecular beam epitaxy process~\cite{wickramasinghe2018high}. The three and four terminal junctions (Fig.~1b,c,d) were fabricated by a combination of electron beam lithography and wet etching (Fig.~1d). In the first step, the (Al)/(quantum well) layers are wet-etched through a resist mask using $\mathrm{H_2O:Ci:H_3PO_4:H_2O_2}~(220:55:3:3)$ for 4-8 min to define multiple fork-like and cross-like geometries, electrically isolated from each other by the mesas. In the second step, Al is removed from the desired junction region using Transene Type D at 50C. The fabrication procedure is similar to that used for two-terminal junctions~\cite{mayer2018superconducting}. The chips were wirebonded to a printed circuit board with built-in discrete-element filters, which was mounted to the Copper probe of a bottom-loading Blue Fors dilution refrigerator. The probe's temperature was between $10-15~\textrm{mK}$ during the experiment. Magnetic field was applied by an external hand-made superconducting coil. Differential resistance measurements were performed using the standard lock-in technique. The MAR resonances in the channel $i,j$ were identified by plotting the lines satisfying $V_{ij} = 2\Delta/n$ ($n = 1,2,...$) on top of the current-biased data (Fig.~2 and Fig.~3e,f) and adjusting $\Delta$ for the best match (Suppl. Mat.).

\noindent \textbf{Theoretical.} We consider a model of a general $N$-terminal Josephson junction (Fig.~1a) with the terminals characterized by the uniform gap $\Delta$ and phases $\phi_j$, $j=1,..., N$ and with the exposed semiconductor junction region characterized by a multi-channel energy-independent scattering matrix $\hat S$. The spectrum of discrete Andreev bound states $E^A_n$, $n = 1,2,...$ of such a junction is given in terms of $\hat S$ through the Beenakker's determinant equation~\cite{beenakker1991universal, van2014single,  xie2017topological} on energy $E$:
\begin{equation}
\textrm{det}[1-\alpha(E/\Delta)\hat{r}\hat{S}^*\hat{r}^*\hat{S}]=0.
\end{equation}
Here $\alpha(x) = \exp(-2i\arccos x)$ defines the energy-dependent phase shift of Andreev reflection and $\hat{r} = \exp (i\hat\phi)$, where $\hat \phi = \mathrm{diag}[\phi_1, \phi_2, ..., \phi_N]$. In the absence of other symmetries, the minimal requirement on $\hat S$ is unitarity, $\hat{S}^{\dagger} = \hat S^{-1}$, imposed by the conservation of normal currents in the junction. In case of a time-reversal symmetry $\hat S$ must be orthogonal. The ground state energy of the junction $E_g$ and the supercurrents $I_j$ ($j = 1, ..., N$) are given by
\begin{equation}
E_g = -(1/2)\sum_n E^A_n;~~
I_j =\partial E_g/\partial\phi_j (2e/\hbar),
\end{equation}
where the sum in the former equation runs over all positive ABS~\cite{de2018superconductivity}. The effect of extended phase-sensitive states can be neglected. Assuming the $j=N$ terminal is grounded, we can set $\phi_N =0$ and measure all other phases with respect to the grounded terminal. The Josephson energy $E_J$ of the multi-terminal junction can be defined as $E_g = E_J(\phi_1, ..., \phi_{N-1})$. It is a $2\pi$-periodic function of $N-1$ variables and the supercurrents conservation $I_N = -\sum_{j=1}^{j=N-1}I_j$ is satisfied automatically.

For $N=3$ devices (Fig.~1b,c), the physically allowed non-dissipative currents are obtained by sweeping with variables $\phi_1$ and $\phi_2$ the surface of a torus. The boundary of the resulting simply-connected shape defines a 2D contour of possible critical current pairs $\{I_1^c, I_2^c\}$. The ``critical current" of a three-terminal junction consists of three such contours obtained by permuting the choice of the grounded electrode. For $N=4$ case (Fig.~1d), we use a pairwise balanced current bias without an explicit grounding. In this case it is more convenient to formally keep a four-variable Josephson energy but impose a constraint  $I_i = -I_j = I_{ij}$ on the currents. Under such symmetric bias conditions, the critical current of a 4-terminal junction is also a set of three 2D contours $\{I_{13}^c, I_{24}^c\}$, $\{I_{12}^c, I_{34}^c\}$, and $\{I_{23}^c, I_{41}^c\}$.

All three critical current contours of the three-terminal device (Fig.~2a,b) can be readily matched by a randomly generated orthogonal matrix with three channels per terminal (Suppl. Mat.). The theory in Fig.~4c,d was produced using a similarly generated four-terminal orthogonal (time-reversal symmetric) and unitary (time-reversal broken) scattering matrices. We have checked that the short-scale fluctuations of the contours -- the key feature of the time-reversal broken systems -- can be reproduced by increasing the number of transport channels from three to ten, which is more appropriate for the experimental device parameters (Suppl. Mat.).


\bibliography{SuperconductingCircuits}

\end{document}